\documentclass[12pt]{article}
\usepackage[T2A]{fontenc}
\usepackage[cp1251]{inputenc}
\usepackage{hyperref}
\usepackage{a4wide}
\usepackage[english]{babel}
\usepackage{amsmath,latexsym,amsthm,amsfonts}
\usepackage{amssymb,amscd}
\usepackage{graphicx}

\begin{document}
\title{ Pricing Currency Derivatives with Markov-modulated
 L$\acute{\textnormal{e}}$vy Dynamics}
\author{Anatoliy Swishchuk$^1$, Maksym Tertychnyi$^2$,  Robert Elliott$^3$}
\date{}
\maketitle
\begin{footnotesize}
\begin{tabbing}
$^1$ \= Department of Mathematics and Statistics, University of Calgary, Canada\\
\> {\em aswish@ucalgary.ca}\\
$^2$ \=  Department of Mathematics and Statistics, University of Calgary, Canada\\
\> {\em mtertych@ucalgary.ca, maksym.tertychnyi@gmail.com}\\
$^3$ \> Department of Mathematics and Statistics, University of Calgary, Canada\\
\>{\em relliott@ucalgary.ca, robert.elliott@haskayne.ucalgary.ca}\\
\end{tabbing}
\end{footnotesize}
\begin{abstract}
Using a L$\acute{\textnormal{e}}$vy  process we  generalize formulas in Bo {\it et al.} (2010) for the  Esscher transform parameters for the log-normal distribution which ensure the martingale condition holds  for the  discounted foreign exchange rate. Using these values of the parameters  we find a  risk-neural measure and provide new formulas for the distribution of jumps, the mean jump size,  and the Poisson process intensity  with respect to to this measure. The formulas for a European call foreign exchange  option  are also derived. We apply these formulas  to the case of the log-double exponential distribution of  jumps. We provide numerical simulations for the  European call foreign exchange option prices with different parameters.
\end{abstract}
\thispagestyle{empty}



\section{\bf Introduction}
Until the early 1990s the existing
academic literature on the pricing of foreign currency options could be divided into two
categories. In the first, both domestic and foreign interest rates were assumed to be constant whereas
the spot exchange rate is assumed to be stochastic. See, e.g., Jarrow et al (1981, \cite{b3}). The second class
of models for pricing foreign currency options incorporated stochastic interest rates, and were based on
Merton's 1973, \cite{b22}) stochastic interest rate model for pricing equity options. See, e.g., Grabbe (1983, \cite{b16}),
Adams et al (1987, \cite{b1}). Unfortunately, this pricing approach did not integrate a full term structure
model into the valuation framework.
To our  knowledge, Amin et al. (1991, \cite{b3}) were the first
 to start discussing and building a general
framework to price contingent claims on foreign currencies under stochastic interest rates using the
Heath et al. (1987) model of term structure. Melino et al. (1991, \cite{b20}) examined the foreign exchange rate
process, (under a deterministic interest rate), underlying  observed option prices and Rumsey (1991, \cite{b28})
considered cross-currency options. Mikkelsen (2001, \cite{b24}) investigated by simulation cross-currency options using market
models of interest rates and deterministic volatilities for spot exchange rates. Schlogl
(2002, \cite{b29}) extended market models to a cross-currency framework. Piterbarg (2005, \cite{b26}) developed a model
for cross-currency derivatives such as PRDC swaps with calibration for currency options; he used
neither market models nor stochastic volatility models. In Garman et al. (1983, \cite{b9}) and Grabbe (1983, \cite{b16}),
foreign exchange option valuation formulas were derived under the assumption that the exchange rate
follows a diffusion process with continuous sample paths.
Takahashi et al. (2006, \cite{b32}) proposed a new approximation formula for the valuation of currency options
 using  jump-diffusion stochastic volatility processes for spot exchange rates in a stochastic interest rates
environment. In particular, they applied the  market models developed by Brace et al. (1998), Jamshidian
(1997, \cite{b15}) and Miltersen et al. (1997, \cite{b23}) to model the term structure of interest rates. Also, Ahn et al. (2007, \cite{b2})
derived explicit formulas for European foreign exchange call and put options values when the
exchange rate dynamics are governed by jump-diffusion processes.
Hamilton (1988) was the first to investigate the term structure of interest rates by rational expectations
econometric analysis of changes in regime. Goutte et al. (2011, \cite{b10}) studied foreign exchange
rates using a modified Cox-Ingersoll-Ross model under a Hamilton Markov regime switching
framework. Zhou et al. (2012, \cite{b33}) considered an accessible implementation of interest rate models with
regime-switching. Siu et al (2008, \cite{b31}) considered pricing currency options under a two-factor Markov modulated
stochastic volatility model.  Swishchuk and Elliott   applied hidden Markov models for pricing options in \cite{b31-1}.  Bo et al. (2010, \cite{b7}) discussed a  Markov-modulated jump-diffusion,
(modeled by a compound Poisson process), for currency option pricing.
We note that currency derivatives for domestic and foreign equity markets and for the exchange rate
between the domestic currency and a fixed foreign currency with constant interest rates were discussed
in Bjork (1998, \cite{b6}). We also mention that currency conversion for forward and swap prices with constant
domestic and foreign interest rates were discussed in Benth et al. (2008, \cite{b5}).

In this article we generalize the results of  \cite{b7} to a case when the dynamics of the  FX rate are driven by a general L$\acute{\textnormal{e}}$vy process (\cite{b25}). The main results of our research are as follows:

1) In section 2 we generalize the formulas of  \cite{b7} for Esscher transform parameters which ensure that the martingale condition for the discounted foreign exchange rate is a martingale for a general L$\acute{\textnormal{e}}$vy process (see \eqref{4.20}). Using these values of the parameters  (see \eqref{4.29}, \eqref{4.30}) we proceed to a risk-neural measure and provide new formulas for the distribution of jumps, (see \eqref{4.26}), the mean jump size (see \eqref{4.10}), and  the Poisson process intensity with respect  to the measure(see \eqref{4.9}).  At the end of section 2   pricing formulas for European call foreign exchange options are given. (They are similar to those in \cite{b7}, but the mean jump size and the Poisson process intensity with respect to the new risk-neutral measure are different).

2) In section 3 we apply formulas \eqref{4.9}, \eqref{4.10}, \eqref{4.29}, \eqref{4.30} to the case of  log double exponential processes, (see \eqref{4.40}), for jumps (see \eqref{4.47}-\eqref{4.50}).

3) In section 5 we provide numerical simulations of European call foreign exchange option prices for  different parameters, (in the case of log-double exponential and exponential distributions of jumps):  $S/K$, where $S$ is the initial spot FX  rate, $K$  is the strike FX rate for a  maturity time $T$ and the parameters $\theta_1, \theta_2$ refer to the log-double exponential distribution.

In the Appendix codes for Matlab  functions used in numerical simulations of option prices are provided.

\section{Currency option pricing  for general L$\acute{\textnormal{e}}$vy processes}

Let $(\Omega, \mathcal{F}, \textbf{P})$ be a complete probability space with a  probability measure $\textbf{P}$. Consider a continuous-time, finite-state
Markov chain $\xi= \{\xi_t\}_{0\leq t \leq T}$  on
$(\Omega, \mathcal{F}, \textbf{P})$ with a state space $\mathcal{S}$, the set of unit vectors $(e_1,\cdots, e_n)\in \mathbb{R}^n$ with a rate matrix $\Pi$\footnote{In our numerical simulations we consider three-state Markov chain and calculate elements in $\Pi$ using Forex market EURO/USD currency pair}.
The dynamics of the chain are given by:
 \begin{equation}\label{3.11-1}
 \xi_t=\xi_0+\int_0^t \Pi \xi_u du+M_t \in \mathbb{R}^n,
\end{equation}
where  $M = \{M_t, t \geq 0\}$ is a $\mathbb{R}^n$-valued martingale with respect to $(\mathcal{F}_t^\xi)_{0 \leq t \leq T}$, the $\textbf{P}$-augmentation of the natural filtration $(\mathcal{F}_t)_{0 \leq t \leq T}$, generated by the Markov chain $\xi$. Consider  a Markov-modulated Merton jump-diffusion which models the dynamics of the spot FX rate, given by the following stochastic differential equation (in the sequel SDE, see \cite{b7}):
\begin{equation}\label{3.11}
dS_t= S_{t_-}\left(\mu_t dt+\sigma_t d W_t+(e^{Z_{t_-}}-1)dN_t\right).
\end{equation}
Here $\mu_t$  is drift parameter; $W_t$ is a  Brownian motion, $\sigma_t$ is the  volatility; $N_t$ is a  Poisson Process with intensity $\lambda_t$,  $e^{Z_{t_-}}-1$ is the amplitude of the jumps, given the  jump arrival time. The distribution of $Z_t$ has a density $\nu(x), x\in \mathbb{R}$.
The parameters $\mu_t$,\;$\sigma_t$,\;$\lambda_t$ are modeled using the finite state Markov chain:
\begin{align}\label{3.11-1}
&\mu_t:=<\mu, \xi_t>, \; \mu \in \mathbb{R}^n_+;\notag \\
&\sigma_t:=<\sigma, \xi_t>, \; \sigma \in \mathbb{R}^n_+;\notag\\
&\lambda_t:=<\lambda, \xi_t>, \; \lambda \in \mathbb{R}^n_+.
\end{align}
The solution of \eqref{3.11} is  $S_t=S_0 e^{L_t}$, (where $S_0$ is the spot FX rate at time $t=0$). Here  $L_t$ is given by the formula:
\begin{equation}\label{3.12}
L_t=\int_0^t(\mu_s-1/2 \sigma_s^2) ds+\int_0^t \sigma_s dW_s+\int_0^t Z_{s_{-}}d N_s.
\end{equation}

There is more than one equivalent martingale measure for this market driven by a Markov-modulated jump-diffusion model.  We shall
define the regime-switching generalized Esscher transform to determine a specific equivalent martingale measure.

Using Ito's formula we can derive a stochastic differential equation for the discounted spot FX rate.
To define the discounted spot FX rate   we need to introduce domestic and foreign riskless interest rates for bonds in the domestic and foreign currency.

The domestic and foreign interest rates $(r^d_t)_{0\leq t\leq T}$,
$(r^f_t)_{0\leq t\leq T}$ are defined using the Markov
chain $(\xi_t)_{0\leq t\leq T}$ (see \cite{b7}):
$$ r^d_t=\langle r^d, \xi_t\rangle, r^d \in \mathbb{R}_+^n,$$
$$ r^f_t=\langle r^f, \xi_t\rangle, r^f \in \mathbb{R}_+^n.$$
The discounted spot FX rate is:
\begin{equation}\label{3.13}
 S^d_t=\textnormal{exp}\left(\int_0^t (r^d_s-r^f_s) ds\right)S_t, \quad
 0\leq t\leq T.
\end{equation}
Using \eqref{3.13}, the differentiation formula, see Elliott et al. (1982, \cite{b11-1}) and the   stochastic differential equation for the spot FX rate \eqref{3.11} we find the stochastic differential equation for the discounted discounted spot FX rate:
\begin{equation}\label{3.14}
d S_{
t_-}^d= S_{t_-}^d(r^d_t-r^f_t+\mu_t) dt+
S_{t_-}^d\sigma_t d W_t
  +S_{t_-}^d(e^{Z_{t_-}}-1) d N_t.
\end{equation}

To derive the main results consider the log spot FX rate

$$Y_t= \log\left(\frac{S_t}{S_0}\right)$$
Using the differentiation formula:

$$Y_t=C_t+J_t,$$
where $C_t, J_t$ are the  continuous and diffusion part of $Y_t$. They are given  in \eqref{3.15}, \eqref{3.16}:

\begin{equation}\label{3.15}
  C_t=\int_0^t\left (r^d_s-r^f_s+\mu_s
  \right)ds+\int_0^t \sigma_s dW_s,
\end{equation}

\begin{equation}\label{3.16}
 J_t=\int_0^t Z_{s_-}dN_s.
\end{equation}

 Let  $(\mathcal{F}_t^Y)_{0 \leq t \leq T}$ denote the $\textbf{P}$-augmentation of the natural filtration $(\mathcal{F}_t)_{0 \leq t \leq T}$, generated by  $Y$. For each $t \in [0, T]$ set $\mathcal{H}_t=\mathcal{F}_t^Y\vee \mathcal{F}_t^\xi$. Let us also define two families of regime switching parameters

$(\theta_s^c)_{0\leq t\leq T}$, $(\theta_s^J)_{0\leq t\leq T}$:
$\theta_t^m=<\theta^m, \xi_t>$, $\theta^m=(\theta_1^m,..., \theta_n^m)\subset \mathbb{R}^n$, $m=\{c, J\}$.

Define a random Esscher transform $\textbf{Q}^{\theta^c, \theta^J}\sim \textbf{P}$ on $\mathcal{H}_t$ using these  families of parameters $(\theta_s^c)_{0\leq t\leq T}$, $(\theta_s^J)_{0\leq t\leq T}$ (see \cite{b7}, \cite{b11}, \cite{b12} for details):

\begin{equation}\label{4.1}
L_t^{\theta^c, \theta^J}=\frac{d \textbf{Q}^{\theta^c, \theta^J}}{d
\textbf{P}}\biggl|_{\mathcal{H}_t}=:
\end{equation}
$$
  \frac{\textnormal{exp} \left(\int_0^t \theta_s^c dC_s +\int_0^t \theta_{s_-}^J d J_s\right )}
  {\mathbb{E}\left [\textnormal{exp} \biggl(\int_0^t \theta_s^c dC_s +\int_0^t \theta_{s_-}^J d J_s\right )\biggl | \mathcal{F}_t^\xi\biggl]}.
$$

The explicit formula for the density $L_t^{\theta^c, \theta^J}$ of the Esscher transform is given in the following Theorem. A similar statement is proven for the log-normal distribution in \cite{b7}. The formula below can be obtained by another approach, considered by Elliott and Osakwe (\cite{b11-2}).

\textbf{Theorem 2.1.} \textit{For $0\leq t \leq T$ the density $L_t^{\theta^c, \theta^J}$ of Esscher transform defined in \eqref{4.1} is given by }
\begin{equation}\label{4.2}
L_t^{\theta^c, \theta^J}=\textnormal{exp}\left(\int_0^t \theta_s^c
\sigma_s dW_s -1/2 \int_0^t (\theta_s^c \sigma_s)^2
ds\right)\times
\end{equation}
$$\textnormal{exp}\left(\int_0^t \theta_{s_-}^J Z_{s_-}d N_s-\int_0^t \lambda_s\biggl(\int_{\mathbb{R}}e^{\theta_s^J x}\nu(dx)-1\right)ds\biggl).$$

\textit{In addition,the random Esscher transform density
$L_t^{\theta^c,\theta^J}$ (see \eqref{4.1}, \eqref{4.2}) is an exponential $(\mathcal{H}_t)_{0\leq t\leq T}$ martingale and satisfies the following SDE:}
\begin{equation}\label{4.3}
  \frac{d L_t^{\theta^c,
 \theta^J}}{L_{t_-}^{\theta^c,
 \theta^J}}=\theta_t^c \sigma_t dW_t+(e^{\theta_{t_-}^J
 Z_{t_-}}-1)d N_t-\lambda_t \left(\int_{\mathbb{R}}e^{\theta_t^J x}\nu(dx)-1\right)dt.
\end{equation}

\textbf{Proof of Theorem 2.1. }
The compound Poisson Process, driving the jumps $\sum_0^{N_t}(e^{Z_{s_-}}-1)$ and the Brownian motion $W_t$ are independent processes. Consequently:
$$\mathbb{E}\left [\textnormal{exp} \biggl(\int_0^t \theta_s^c dC_s +\int_0^t \theta_{s_-}^J d J_s\right )\biggl | \mathcal{F}_t^\xi\biggl]=$$

\begin{equation}\label{4.4}\mathbb{E}\left [\exp \biggl(\int_0^t \theta_s^c (\mu_s-1/2\sigma_s^2)ds +\int_0^t \theta_s^c \sigma_s dW_s)\biggl) \biggl|\mathcal{F}_t^\xi\right]
 \mathbb{E}\left [\textnormal{exp}\biggl(\int_0^t \theta_{s_-}^J Z_{s_-} d N_s\biggl) \biggl| \mathcal{F}_t^\xi\right]. \end{equation}
Let us calculate:
 $$\mathbb{E}\left [\textnormal{exp}\biggl(\int_0^t \theta_{s_-}^J Z_{s_-} d N_s\biggl) \biggl| \mathcal{F}_t^\xi\right].$$
Write $$\Gamma_t:=\exp\left(\int_0^t\alpha_{s_-}dN_s \right), \alpha_s=\theta_s^J Z_s.$$
Using the differentiation rule (see \cite{b11-1}) we obtain the following representation of $\Gamma_t$:
\begin{equation}\label{4.4-1}
\Gamma_t=\Gamma_0+M_t^J+\int_{]0,t]}\Gamma_s\int_{\mathbb{R}}(e^{\alpha_s}-1)\nu(dx)\lambda_s ds,
\end{equation}
where $$M_t^J=\int_{]0, t]}\Gamma_{s_-}(e^{\alpha_s}-1)dNs- \int_{]0,t]}\Gamma_s\int_{\mathbb{R}}(e^{\alpha_s}-1)\nu(dx)\lambda_s ds$$
is a martingale with respect to $\mathcal{F}_t^\xi$.
Using this fact and \eqref{4.4-1} we obtain:
\begin{equation}\label{4.5}
\mathbb{E}\left [\textnormal{exp}\biggl(\int_0^t \theta_{s}^J Z_{s_-}dN_s\biggl) \biggl| \mathcal{F}_t^\xi\right]=\exp\biggl(\int_0^t \lambda_s\biggl(\int_{\mathbb{R} }e^{\theta_s^J x}\nu(dx)-1 \biggl) ds\biggl)
\end{equation}
We have from the differentiation rule:
\begin{equation}\label{4.5-1}
\mathbb{E}\left[e^ {u\int_0^t\sigma_s d W_s}\right]=
\exp \left \{\frac{1}{2}u^2\int_0^t \sigma_s^2 ds\right\},
\end{equation}
 where $\sigma_t$  is the volatility of a market.

Substituting \eqref{4.5} and \eqref{4.5-1} into \eqref{4.4} we obtain:
$$
\mathbb{E}\left [\textnormal{exp} \biggl(\int_0^t \theta_s^c dC_s +\int_0^t \theta_{s_-}^J d J_s\right )\biggl | \mathcal{F}_t^\xi\biggl]=
$$

\begin{equation}\label{4.6}
\exp \biggl(\int_0^t \theta_s^c (\mu_s-1/2\sigma_s^2)ds +\frac{1}{2}\int_0^t (\theta_s^c \sigma_s)^2 ds )\biggl)
\exp\biggl(\int_0^t \lambda_s\biggl(\int_{\mathbb{R} }e^{\theta_s^J x}\nu(dx)-1 \biggl) ds\biggl).
 \end{equation}
Substituting \eqref{4.6} into the expression for $L_t^{\theta^c,\theta^J}$ in \eqref{4.1} we obtain:

\begin{equation}\label{4.7}
L_t^{\theta^c,\theta^J}=\exp \biggl(\int_0^t \theta_s^c (\mu_s-1/2\sigma_s^2)ds +\int_0^t \theta_s^c \sigma_s dW_s)\biggl)\textnormal{exp}\biggl(\int_0^t \theta_{s_-}^J Z_{s_-} d N_s\biggl)\times
\end{equation}
$$
\biggl[\exp \biggl(\int_0^t \theta_s^c (\mu_s-1/2\sigma_s^2)ds +\frac{1}{2}\int_0^t (\theta_s^c \sigma_s)^2 ds )\biggl)
\exp\biggl(\int_0^t \lambda_s\biggl(\int_{\mathbb{R} }e^{\theta_s^J x}\nu(dx)-1 \biggl) ds\biggl)\biggl]^{-1}=
$$
$$
\textnormal{exp}\left(\int_0^t \theta_s^c
\sigma_s dW_s -1/2 \int_0^t (\theta_s^c \sigma_s)^2
ds\right)\times
$$
$$\textnormal{exp}\left(\int_0^t \theta_{s_-}^J Z_{s_-}d N_s-\int_0^t \lambda_s\biggl(\int_{\mathbb{R}}e^{\theta_s^J x}\nu(dx)-1\right)ds\biggl).
$$

If we represent $L_t^{\theta^c,\theta^J}$ in the form  $L_t^{\theta^c,\theta^J}=e^{X_t}$ (see \eqref{4.7}) and apply differentiation rule we obtain \eqref{4.3}. It follows from \eqref{4.3} that $L_t^{\theta^c,\theta^J}$ is a martingale.   $\square$

We shall derive the following condition for the discounted spot FX rate (\eqref{3.13}) to be martingale. These conditions will be used to calculate the risk-neutral  Esscher transform parameters $(\theta_s^{c, \ast})_{0\leq t\leq T}$, $(\theta_s^{J, \ast})_{0\leq t\leq T}$ and give to the measure $\textbf{Q}$. Then we shall use these values to find the no-arbitrage price of European call currency derivatives.

\textbf{Theorem 2.2.} \textit{Let the random Esscher transform be defined by \eqref{4.1}. Then
the martingale condition(for $S^d_t$, see \eqref{3.13}) holds if and only if the Markov
modulated parameters ($\theta_t^c, \theta_t^J, 0\leq t\leq T$)
satisfy for all $0\leq t\leq T$ the condition:}
\begin{equation}\label{4.8}
r^f_t-r^d_t+\mu_t+\theta_t^c\sigma_t^2+\lambda_t^{\theta,
J }k_t^{\theta, J}=0.
\end{equation}
\textit{Here the random Esscher transform intensity $\lambda_t^{\theta,
J}$ of the Poisson Process and the main percentage jump size
$k_t^{\theta, J}$ are, respectively, given by}
\begin{equation}\label{4.9}
\lambda_t^{\theta, J}=\lambda_t \int_{\mathbb{R}}e^{\theta_s^J x}\nu(dx),
\end{equation}
\begin{equation}\label{4.10}
 k_t^{\theta, J}=\frac{\int_{\mathbb{R}}e^{(\theta_t^J +1)x}\nu(dx)}{\int_{\mathbb{R}}e^{\theta_t^J x}\nu(dx)}-1,
\end{equation}
\textit{as long as} $\int_{\mathbb{R}}e^{\theta_t^J x}\nu(dx)<+\infty$, $\int_{\mathbb{R}}e^{(\theta_t^J +1) x}\nu(dx)<+\infty$.

\textbf{Proof of Theorem 2.2.}
The martingale condition for the discounted spot FX rate $S^d_t$ is
\begin{equation}\label{4.11}
\mathbb{E}^{\theta^c, \theta^J}[S^d_t| \mathcal{H}_u]=S^d_u, \quad
t\geq u.
\end{equation}
To derive this condition a Bayes' formula is used:
\begin{equation}\label{4.12}
  \mathbb{E}^{\theta^c,
\theta^J}[S^d_t|\mathcal{H}_u]=\frac{\mathbb{E}[L_t^{\theta^c,
\theta^J} S^d_t|\mathcal{H}_u]}{\mathbb{E}[L_t^{\theta^c,
\theta^J} |\mathcal{H}_u]},
\end{equation}
 taking into account that $L_t^{\theta^c,\theta^J}$ is a martingale with respect to $\mathcal{H}_u$, so:
\begin{equation}\label{4.13}
\mathbb{E}\biggl[L_t^{\theta^c, \theta^J} \biggl | \mathcal{H}_u\biggl]=L_u^{\theta^c,
\theta^J}.
\end{equation}

Using  formula \eqref{3.13} for the solution of the  SDE for the spot FX rate, we obtain an expression for the discounted  spot FX rate in the following form:

\begin{equation}\label{4.14}
S_t^d=S_u^d \exp\biggl(\int_u^t(r_s^f-r_s^d+\mu_s-1/2 \sigma_s^2) ds+\int_u^t \sigma_s dW_s+ \int_u^t Z_{s_-} d N_s\biggl), \quad t\geq u.
\end{equation}
Then, using \eqref{4.2}, \eqref{4.14} we can rewrite\eqref{4.13} as:
\begin{equation}\label{4.15}
\mathbb{E}\biggl[\frac{L_t^{\theta^c, \theta^J}}{L_u^{\theta^c,
\theta^J}}S^d_t \biggl | \mathcal{H}_u\biggl]=S_u^d \;\mathbb{E}\biggl[\exp \left(\int_u^t \theta_s^c
\sigma_s dW_s -1/2 \int_0^t (\theta_s^c \sigma_s)^2
ds\right)\times
\end{equation}
$$\textnormal{exp}\left(\int_u^t \theta_{s_-}^J Z_{s_-}d N_s-\int_u^t \lambda_s\biggl(\int_{\mathbb{R}}e^{\theta_s^J x}\nu(dx)-1\right)ds\biggl)\times $$
$$
\exp\biggl(\int_u^t(r_s^f-r_s^d+\mu_s-1/2 \sigma_s^2) ds+\int_u^t \sigma_s dW_s+ \int_u^t Z_{s_-} d N_s\biggl)| \mathcal{H}_u\biggl]=
$$
\begin{equation}\label{4.16}
 S_u^d \;\mathbb{E}\biggl[\exp \left(\int_u^t (\theta_s^c+1)
\sigma_s dW_s -1/2 \int_u^t ((\theta_s^c+1) \sigma_s)^2
ds\right)\times
\end{equation}
$$
\exp\biggl(\int_u^t(r_s^F-r_s^D+\mu_s+  \theta_s^c\sigma_s^2) ds\biggl)\;
\exp \biggl( \int_u^t \lambda_s\bigl(\int_{\mathbb{R}}e^{\theta_s^J x}\nu(dx)-1\bigl)ds\biggl)|\mathcal{H}_u\biggl]\times
$$
$$
\mathbb{E}\biggl[\exp \left(\int_u^t(\theta_s^c+1)Z_{s_-}d N_s\right)| \mathcal{H}_u\biggl].
$$

Using the expression for the characteristic function of Brownian motion (see \eqref{4.5-1}) we obtain:
\begin{equation}\label{4.17}
 \mathbb{E}\biggl[\exp \left(\int_u^t (\theta_s^c+1)
\sigma_s dW_s -1/2 \int_u^t ((\theta_s^c+1) \sigma_s)^2
ds\right)| \mathcal{H}_u\biggl]=1.
\end{equation}

From \eqref{4.5} we obtain:
\begin{equation}\label{4.18}
\mathbb{E}\biggl[\exp \left(\int_u^t(\theta_s^c+1)Z_{s_-}d N_s\right)| \mathcal{H}_u\biggl]=\exp\biggl(\int_0^t \lambda_s\biggl(\int_{\mathbb{R} }e^{(\theta_s^J+1) x}\nu(dx)-1 \biggl) ds\biggl).
\end{equation}
Substituting \eqref{4.17}, \eqref{4.18} into \eqref{4.16} we obtain finally:

\begin{equation}\label{4.19}
\mathbb{E}\biggl[\frac{L_t^{\theta^c, \theta^J}}{L_u^{\theta^c,
\theta^J}}S^d_t \biggl | \mathcal{H}_u\biggl]=S_u^d \exp\biggl(\int_u^t(r_s^f-r_s^d+\mu_s+  \theta_s^c\sigma_s^2) ds\biggl)\times
\end{equation}
$$
\exp \biggl( -\int_u^t \lambda_s\bigl(\int_{\mathbb{R}}e^{\theta_s^J x}\nu(dx)-1\bigl)ds\biggl)\;\exp \biggl( \int_u^t \lambda_s\bigl(\int_{\mathbb{R}}e^{(\theta_s^J+1) x}\nu(dx)-1\bigl)ds\biggl).
$$

From \eqref{4.19} we obtain the  martingale condition for discounted spot FX rate:

\begin{equation}\label{4.20}
r_t^f-r_t^d+\mu_t+  \theta_t^c\sigma_t^2+\lambda_t \biggl[\int_{\mathbb{R}}e^{(\theta_s^J+1) x}\nu(dx)-\int_{\mathbb{R}}e^{\theta_s^J x}\nu(dx)\biggl]=0.
\end{equation}
We now prove, that under the Esscher transform the new Poisson process intensity and mean jump size are given by \eqref{4.9}, \eqref{4.10}.

Note that  $L_t^J=\int_0^t Z_{s_-}dN_s$ is the jump part of L$\acute{\textnormal{e}}$vy process in the formula \eqref{3.12} for the solution of SDE for spot FX rate. We have:
\begin{equation}\label{4.21}
   \mathbb{E}_{\textbf{Q}} \left[ e^{L_t^J}\right]=\int_{\Omega}\exp\left(\int_0^t Z_{s_-}dN_s\right)L_t^{\theta^{c,\ast},\theta^{J, \ast}}(\omega)dP(\omega),
\end{equation}
where $\textbf{P}$ is the initial  probability measure and  $\textbf{Q}$ is the new risk-neutral measure.
Substituting the density of Esscher transform \eqref{4.2} into \eqref{4.21} we have (see also \cite{b11-2}):
\begin{equation}\label{4.22}
 \mathbb{E}_\textbf{Q} \left[ e^{L_t^J}\right]=\mathbb{E}_\textbf{P}\biggl[\textnormal{exp}\left(\int_0^t \theta_s^c
\sigma_s dW_s -1/2 \int_0^t (\theta_s^c \sigma_s)^2
ds\right)-
\end{equation}
$$\int_0^t \lambda_s\biggl(\int_{\mathbb{R}}e^{\theta_s^J x}\nu(dx)-1\biggl)ds\biggl) \biggl] \;\mathbb{E}_\textbf{P}\biggl[\textnormal{exp}\left(\int_0^t (\theta_{s}^J+1) Z_{s_-}d N_s\right)\biggl].
$$
Using \eqref{4.5} we obtain:
\begin{equation}\label{4.23}
\mathbb{E}_\textbf{P}\biggl[\textnormal{exp}\left(\int_0^t (\theta_{s}^J+1) Z_{s_-}d N_s\right)\biggl]=\exp\biggl(\int_0^t \lambda_s\biggl(\int_{\mathbb{R} }e^{(\theta_s^J +1)x}\nu(dx)-1 \biggl) ds\biggl)
\end{equation}
Substitute \eqref{4.23} into \eqref{4.22} and taking into account the  characteristic function of Brownian motion (see \eqref{4.5-1}) we obtain:
\begin{equation}\label{4.24}
 \mathbb{E}_\textbf{Q} \left[ e^{L_t^J}\right]=\exp\biggl(\int_0^t \lambda_s\biggl(\int_{\mathbb{R} }e^{\theta_s^J x}\nu(dx)\left[\frac{\int_{\mathbb{R}}e^{(\theta_s^J +1)x}\nu(dx)}{\int_{\mathbb{R}}e^{\theta_s^J x}\nu(dx)}-1\right] \biggl) ds\biggl).
\end{equation}

Returning to the initial measure $\textbf{P}$, but with different $\lambda_t^{\theta, J}, k_t^{\theta, J}$,we have:
\begin{equation}\label{4.25}
 \mathbb{E}_{\tilde{\lambda}, \tilde{\nu}} \left[ e^{L_t^J}\right]=\exp\biggl(\int_0^t \lambda_s^{\theta, J}\biggl(\int_{\mathbb{R}}e^x \tilde{\nu}(dx)-1\biggl)ds\biggl).
\end{equation}
Formula \eqref{4.9} for the  new intensity $\lambda_t^{\theta, J}$ of Poisson process follows directly from \eqref{4.24}, \eqref{4.25}.
The new density of jumps $\tilde{\nu}$ is defined from \eqref{4.25} by the following formula:
\begin{equation}\label{4.26}
\frac{\int_{\mathbb{R}}e^{(\theta_t^J +1)x}\nu(dx)}{\int_{\mathbb{R}}e^{\theta_t^J x}\nu(dx)}=\int_{\mathbb{R}}e^x \tilde{\nu}(dx).
\end{equation}
We now calculate the  new mean jump size given the jump arrival with respect to the new measure $\textbf{Q}$:
\begin{equation}\label{4.27}
k_t^{\theta, J}=\int_\Omega (e^{Z(\omega)}-1)d\tilde{\nu}(\omega)=\int_{\mathbb{R}} (e^{x}-1)\tilde{\nu}(dx)=\int_{\mathbb{R}} e^{x}\tilde{\nu}(dx)-1=\frac{\int_{\mathbb{R}}e^{(\theta_t^J +1)x}\nu(dx)}{\int_{\mathbb{R}}e^{\theta_t^J x}\nu(dx)}-1,
\end{equation}
where the new measure $\tilde{\nu}(dx)$ is defined by the formula \eqref{4.26}.

We can rewrite the martingale condition \eqref{4.20} for the discounted spot FX rate in the following form:
\begin{equation}\label{4.28}
r_t^f-r_t^d+\mu_t+  \theta_t^c\sigma_t^2+\lambda_t^{\theta, J} k_t^{\theta, J} =0,
\end{equation}
where $\lambda_t^{\theta, J}, k_t^{\theta, J}$  are given by \eqref{4.9}, \eqref{4.10} respectively. $\square$

If we put $  k_t^{\theta, J}=0$, we obtain  the following formulas for the regime switching Esscher transform parameters yielding the martingale condition \eqref{4.28}:
\begin{equation}\label{4.29}
 \theta_t^{c, \ast}=\frac{r_t^d-r_t^f-\mu_t}{\sigma_t^2},
\end{equation}
\begin{equation}\label{4.30}
 \theta_t^{J, \ast}: \frac{\int_{\mathbb{R}}e^{(\theta_t^{J, \ast} +1)x}\nu(dx)}{\int_{\mathbb{R}}e^{\theta_t^{J, \ast} x}\nu(dx)}=1.
\end{equation}
In the next section we shall apply these formulas \eqref{4.29}, \eqref{4.30} to the log-double exponential distribution of jumps.

We now proceed to the general formulas for European calls (see \cite{b7}, \cite{b21}).
For the European call currency options with a strike price $K$ and the time of
expiration $T$ the price at time zero is given by:

\begin{equation}\label{4.31}
  \Pi_0(S, K, T, \xi)=\mathbb{E}^{\theta^{c,\ast}, \theta^{J,
  \ast}}\left[e^{-\int_0^T (r^D_s-r^F_s)ds}(S_T-K)^+\mid \mathcal{F}_t^\xi\right].
\end{equation}

 Let $J_i(t,T)$ denote the occupation time of $\xi$ in state $e_i$ over the period $[t, T], t<T$.  We  introduce several new quantities that will be used in future calculations:
\begin{equation}\label{4.32}
R_{t, T}=\frac{1}{T-t}\int_0^T (r_s^d-r_s^f)ds=\frac{1}{T-t}\sum_{i=1}^n (r_i^d-r_i^f)J_i(t, T),
\end{equation}
where $J_i(t, T):=\int_t^T <\xi_{s},\; e_i>ds$;
\begin{equation}\label{4.33}
U_{t, T}=\frac{1}{T-t}\int_t^T \sigma_s^2 ds=\frac{1}{T-t} \sum_{i=1}^n \sigma_i^2J_i(t, T);
\end{equation}
\begin{equation}\label{4.34}
 \lambda_{t, T}^{\theta^\ast J}=\frac{1}{T-t} \sum_{i=1}^n \lambda_i^{\theta^\ast J}J_i(t, T);
\end{equation}
\begin{equation}\label{4.35}
\lambda_{t, T}^{\theta^\ast }=\frac{1}{T-t}\int_t^T(1+k_s^{\theta^\ast J})\lambda_s^{\theta^\ast J}ds=\frac{1}{T-t}\sum_{i=1}^n (1+k_i^{\theta^\ast J})\lambda_i^{\theta^\ast J} J_i(t, T);
\end{equation}
\begin{equation}\label{4.36}
  V_{t, T, m}^2=U_{t, T}+\frac{m\sigma_J^2}{T-t},
\end{equation}
where $\sigma_J^2$ is the  variance of the distribution of the jumps.
\begin{equation}\label{4.37}
 R_{t, T, m}=R_{t,T}-\frac{1}{T-t}\int_t^T \lambda_s^{\theta^\ast J}k_s^{\theta^\ast J}ds+\frac{1}{T-t}\int_0^T\frac{\log(1+k_s^{\theta^\ast J})}{T-t}ds=
\end{equation}
$$
R_{t, T}-\frac{1}{T-t}\sum_{i=1}^n \lambda_i^{\theta^\ast J}k_i^{\theta^\ast J}+
\frac{m}{T-t}\sum_{i=1}^n \frac{\log(1+k_i^{\theta^\ast J})}{T-t}J_i(t,T),
$$
where $m$ is the number of jumps in the interval $[t, T]$, $n$ is the number of states of the Markov chain $\xi$.

Note, that in our considerations all these general formulas \eqref{4.32}-\eqref{4.37} mentioned above are simplified by the fact that: $k_t^{\theta^\ast J}=0$ with respect to the new risk-neutral measure $\textbf{Q}$ with Esscher transform parameters given by \eqref{4.29}, \eqref{4.30}.
From the pricing formula in Merton (1976, \cite{b21}) let us define (see \cite{b7})
\begin{equation}\label{4.38}
  \overline{\Pi_0}(S, K,T; R_{0, T}, U_{0, T}, \lambda_{0, T}^{\theta^{\ast}})=\sum_{m=0}^{\infty}\frac{e^{-T\lambda_{0, T}^{\theta^\ast,
  J}}(T\lambda_{0, T}^{\theta^\ast})^m}{m!}\times
\end{equation}
$$BS_0(S, K, T, V^2_{0, T, m}, R_{0, T,
  m})$$
where $BS_0(S, K, T, V^2_{0, T, m}, R_{0, T,
  m})$ is the standard Black-Scholes price formula (see \cite{b6})
with initial spot FX rate $S$, strike price $K$, risk-free rate $r$,
volatility square $\sigma^2$ and time $T$ to maturity.

Then, the European style call option pricing formula takes the form (see \cite{b7}):
\begin{equation}\label{4.39}
\Pi_0(S, K, T)=\int_{[0, t]^n}\overline{\Pi_0}(S, K,T; R_{0, T},
U_{0, T}, \lambda_{0, T}^{\Theta^{\ast,
  J}})\times
\end{equation}
$$\psi(J_1, J_2,..., J_n)  dJ_1...dJ_n,$$
where $\psi(J_1, J_2,..., J_n)$ is the joint probability distribution
density for the occupation time, which is determined by the following characteristic function (See \cite{b11-2}):
\begin{equation}\label{4.39-1}
\mathbb{E} \left[\exp \bigl\{\langle u, J(t,T)\rangle \bigl\}\right]=\langle\exp \{(\Pi + diag(u)) (T-t)\}\cdot\mathbb{E}[\xi_0], \mathfrak{1}\rangle,
\end{equation}
where $\mathfrak{1}\in \mathbb{R}^n$ is a vector of ones, $u=(u_1,..., u_n)$ is a vector of transform variables, $J(t,T):=\{J_1(t,T),..., J_n(t,T)\}$.

\section{Currency option pricing  for log-double exponential processes}
The log-double exponential distribution for $Z_{t_-}$ ($e^{Z_{t_-}}-1$ are the jumps in \eqref{3.11}), plays a fundamental role in mathematical finance,  describing the spot FX rate movements over long period of time. It is defined by the following formula of the density function:
\begin{equation}\label{4.40}
 \nu(x)=p\theta_1 e^{-\theta_1 x}\biggl|_{x \geq 0}+(1-p)\theta_2 e^{\theta_2 x}\biggl|_{x < 0}.
\end{equation}
The mean value of this distribution is:
\begin{equation}\label{4.41}
  \textnormal{mean}(\theta_1, \theta_2, p)=\frac{p}{\theta_1}-\frac{1-p}{\theta_2}.
\end{equation}
The variance of this distribution is:
\begin{equation}\label{4.42}
\textnormal{var}(\theta_1, \theta_2, p)=\frac{2p}{\theta_1^2}+\frac{2(1-p)}{\theta_2^2}-\biggl(\frac{p}{\theta_1}-\frac{1-p}{\theta_2}\biggl)^2.
\end{equation}
The double-exponential distribution was first investigated by Kou in \cite{b17}. He also gave economic reasons to use such a type of distribution in Mathematical Finance.
The double exponential distribution has two interesting properties:
the leptokurtic feature (see \cite{b15-1},$\S 3$; \cite{b17} ) and the memoryless property(the probability distribution of X  is memoryless if for any non-negative real numbers $t$ and $s$, we have $\textnormal{Pr}(X>t+s|X>t)=\textnormal{Pr}(X>s)$, see for example \cite{b36}). The last property is inherited from the exponential distribution.

A statistical distribution  has  the leptokurtic feature if there is a  higher peak (higher kurtosis) than  in a normal distribution. This high peak and the corresponding fat tails mean the distribution is more concentrated  around the mean than  a normal distribution, and it has a smaller standard deviation. See details for fat-tail distributions and their applications to Mathematical Finance in \cite{b35}. A leptokurtic distribution means that small changes have less probability because historical values are centered by the mean. However, this also means that large fluctuations have greater probabilities within the fat tails.

\begin{figure}[htb]
\begin{center}
\includegraphics[ scale=0.5]{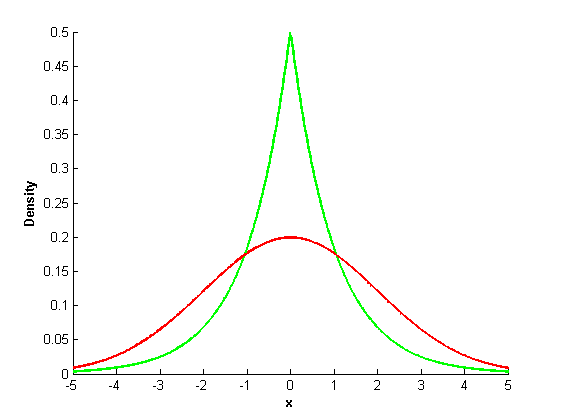}
\end{center}
\caption{Double-exponential distribution (green) vs. normal distribution (red), mean=0, dev=2}
\end{figure}

In \cite{b19} this distribution is applied to  model asset pricing under fuzzy environments. Several advantages of models including log-double exponential distributed jumps over other models described by L$\acute{\textnormal{e}}$vy processes (see \cite{b17}) include:

1) The model describes  properly  some important
empirical results from stock and foreign currency markets. The double exponential
jump-diffusion model is able to reflect the leptokurtic
feature of the return distribution. Moreover, the empirical tests performed
in \cite{b27-1}  suggest that the log-double
exponential jump-diffusion model fits stock data better
than the log-normal jump-diffusion model.

2) The model gives explicit solutions for convenience of computations.

3) The model has an economical interpretation.

4) It has been suggested from
extensive empirical studies that markets tend to have both overreaction and underreaction to  good or bad news.The jump part of the model can be considered as the market response
to outside news. In the absence of
outside news the asset price (or spot FX rate) changes in time as a geometric
Brownian motion. Good or bad news(outer strikes for the FX market in our case) arrive
according to a Poisson process, and the spot FX rate
changes in response according to the jump size distribution.
Since the double exponential distribution
has both a high peak and heavy tails, it can be
applied to model both the overreaction (describing the
heavy tails) and underreaction (describing the high
peak) to outside news.

5) The log double exponential model is   self-consistent. In
Mathematical Finance, it means that a model is  arbitrage-free.

The family of regime switching Esscher transform parameters is defined by \eqref{4.29}, \eqref{4.30}. Let us define $ \theta_t^{J, \ast}$, (the first parameter $\theta_t^{c, \ast}$ has the same formula as in general case) by:
\begin{equation}\label{4.43}
    \int_{\mathbb{R}}e^{(\theta_s^J +1)x}\biggl(p\theta_1 e^{-\theta_1 x}\biggl|_{x \geq 0}+(1-p)\theta_2 e^{\theta_2 x}\biggl|_{x < 0}\biggl) dx=
\end{equation}
$$
\int_{\mathbb{R}}e^{\theta_s^J x}\biggl(p\theta_1 e^{-\theta_1 x}\biggl|_{x \geq 0}+(1-p)\theta_2 e^{\theta_2 x}\biggl|_{x < 0}\biggl) dx
$$
We require an  additional restriction for the convergence of the  integrals in \eqref{4.43}:
\begin{equation}\label{4.44}
   -\theta_2<\theta_t^J<\theta_1.
\end{equation}
Then \eqref{4.43} can be rewritten in the following form:
\begin{equation}\label{4.45}
   \frac{p\theta_1}{\theta_1-\theta_t^J-1}+\frac{(1-p)\theta_2}{\theta_2+\theta_t^J+1}= \frac{p\theta_1}{\theta_1-\theta_t^J}+\frac{(1-p)\theta_2}{\theta_2+\theta_t^J}
\end{equation}
Solving \eqref{4.45} we arrive at the quadratic equation:
\begin{equation}\label{4.46}
   (\theta_t^J)^2(p \theta_1-(1-p)\theta_2)+\theta_t^J(p \theta_1+2\theta_1\theta_2-(1-p)\theta_2)+
\end{equation}
$$
p\theta_1\theta_2^2+p\theta_2\theta_1^2-\theta_2\theta_1^2+\theta_1\theta_2=0.
$$

If $p \theta_1-(1-p)\theta_2\neq0$ we have two solutions and one of them satisfies restriction \eqref{4.44}:
\begin{equation}\label{4.47}
   \theta_t^J=-\frac{p \theta_1+2\theta_1\theta_2-(1-p)\theta_2}{2(p \theta_1-(1-p)\theta_2)}\pm
\end{equation}

$$\frac{\sqrt{(p \theta_1+2\theta_1\theta_2-(1-p)\theta_2)^2-4(p \theta_1-(1-p)\theta_2)(p\theta_1\theta_2^2+p\theta_2\theta_1^2-\theta_2\theta_1^2+\theta_1\theta_2)}}{2(p \theta_1-(1-p)\theta_2)}$$
Then the  Poisson process intensity (see \eqref{4.9}) is:
\begin{equation}\label{4.47-1}
\lambda_t^{\theta, J}=\lambda_t\left(\frac{p\theta_1}{\theta_1- \theta_t^J}+\frac{(1-p)\theta_2}{\theta_2+ \theta_t^J}\right).
\end{equation}
The new mean jump size (see \eqref{4.10}) is:
\begin{equation}\label{4.47-2}
k_t^{\theta, J}=0
\end{equation}
as in the general case.

When we proceed to a new risk-neutral measure $Q$ we have a new density of jumps $\nu$
\begin{equation}\label{4.48}
 \tilde{\nu}(x)=\tilde{p}\theta_1 e^{-\theta_1 x}\biggl|_{x \geq 0}+(1-\tilde{p})\theta_2 e^{\theta_2 x}\biggl|_{x < 0}.
\end{equation}
The new probability $\tilde{p}$ can be calculated using \eqref{4.26}:
\begin{equation}\label{4.49}
  \frac{\frac{p\theta_1}{\theta_1-\theta_t^J-1}+\frac{(1-p)\theta_2}{\theta_2+\theta_t^J+1}}{\frac{p\theta_1}{\theta_1-\theta_t^J}+\frac{(1-p)\theta_2}{\theta_2+\theta_t^J}}=\frac{\tilde{p}\theta_1}{\theta_1-1}+\frac{(1-\tilde{p}\theta_2)}{\theta_2+1}.
\end{equation}
From \eqref{4.49} we obtain an  explicit formula for $\tilde{p}$:

\begin{equation}\label{4.50}
\tilde{p}=\frac{ \frac{\frac{p\theta_1}{\theta_1-\theta_t^J-1}+\frac{(1-p)\theta_2}{\theta_2+\theta_t^J+1}}{\frac{p\theta_1}{\theta_1-\theta_t^J}+\frac{(1-p)\theta_2}{\theta_2+\theta_t^J}}-\frac{\theta_2}{\theta_2+1}}{\frac{\theta_1}{\theta_1-1}-\frac{\theta_2}{\theta_2+1}}.
\end{equation}

\section{Currency option pricing  for log-normal  processes}
Log-normal distribution of jumps with $\mu_J$ the mean, $\sigma_J$ the deviation   (see \cite{b37}), and its applications to currency option pricing was investigated in \cite{b7}. More details of these distributions and other distributions applicable for the Forex market can be found in \cite{b38}. We give here a sketch of results from \cite{b7} to compare them with the case of the  log-double exponential distribution of jumps discussed in this article.  The main goal  of our paper is a generalization of this result for arbitrary L$\acute{\textnormal{e}}$vy processes. The results,\; provided in \cite{b7} are as follows:

\textbf{Theorem 2.3.} \textit{For $0\leq t \leq T$ the density $L_t^{\theta^c, \theta^J}$ of the Esscher transform defined in \eqref{4.1} is given by }
\begin{equation}\label{4.51}
L_t^{\theta^c, \theta^J}=\textnormal{exp}\left(\int_0^t \theta_s^c
\sigma_s dW_s -1/2 \int_0^t (\theta_s^c \sigma_s)^2
ds\right)\times
\end{equation}
$$\textnormal{exp}\left(\int_0^t \theta_{s_-}^J Z_{s_-}d N_s-\int_0^t \lambda_s\biggl(e^{\theta_s^J \mu_J+1/2(\theta_s^J \sigma_J)^2}-1\right)ds\biggl),$$
\textit{where $\mu_J, \sigma_J $ are the mean value and deviation of jumps, respectively.
In addition, the random Esscher transform density
$L_t^{\theta^c,\theta^J}$, (see \eqref{4.1}, \eqref{4.2}), is an exponential $(\mathcal{H}_t)_{0\leq t\leq T}$ martingale and satisfies the following SDE}
\begin{equation}\label{4.52}
  \frac{d L_t^{\theta^c,
 \theta^J}}{L_{t_-}^{\theta^c,
 \theta^J}}=\theta_t^c \sigma_t dW_t+(e^{\theta_{t_-}^J
 Z_{t_-}}-1)d N_t-\lambda_t \left(e^{\theta_t^J \mu_J+1/2(\theta_t^J \sigma_J)^2}-1\right)dt.
\end{equation}

\textbf{Theorem 2.4.} \textit{Let the random Esscher transform be defined by \eqref{4.1}. Then
the martingale condition(for $S^d_t$, see \eqref{3.13}) holds if and only if the Markov
modulated parameters ($\theta_t^c, \theta_t^J, 0\leq t\leq T$)
satisfy for all $0\leq t\leq T$ the condition:}
\begin{equation}\label{4.53}
r^f_i-r^d_i+\mu_i+\theta_i^c\sigma_i^2+\lambda_i^{\theta,
J }k_i^{\theta, J}=0 \;\textnormal{for\; all}\; i, \;1 \leq i \leq \Lambda.
\end{equation}
\textit{where the random Esscher transform intensity $\lambda_i^{\theta,
J}$ of the Poisson Process and the mean percentage jump size
$k_i^{\theta, J}$ are respectively given by}
\begin{equation}\label{4.54}
\lambda_i^{\theta, J}=\lambda_i e^{\theta_i^J \mu_J+1/2(\theta_i^J \sigma_J)^2},
\end{equation}
\begin{equation}\label{4.55}
 k_i^{\theta, J}=e^{\mu_J+1/2\sigma_J^2+\theta_i^J\sigma_J^2}-1 \;\textnormal{for\; all} \;i.
\end{equation}
\textit{The regime switching parameters satisfying the martingale condition \eqref{4.53} are given by the following formulas:
$\theta_i^{c,\ast}$ is the same as in \eqref{4.29},}
\begin{equation}\label{4.56}
 \theta_i^{J,\ast}=-\frac{\mu_J+1/2\sigma_J^2}{\sigma_J^2} \; \textnormal{for\; all}\; i.
\end{equation}
With such a value of a parameter $\theta_i^{J,\ast}$:

\begin{equation}\label{4.561}
  k_i^{\theta^\ast, J}=0, \quad \lambda_i^{\theta^\ast, J}=\lambda_i\left(-\frac{\mu_J}{2\sigma_J^2}+\frac{\sigma_J^2}{8}\right)\; \textnormal{for\; all} \;i.
\end{equation}

Note, that these formulas \eqref{4.54}-\eqref{4.561} follow directly  from our formulas for the case of general L$\acute{\textnormal{e}}$vy process, (see \eqref{4.9},\eqref{4.10},  \eqref{4.30}).  In particular, the fact that $k_i^{\theta^\ast, J}=0$ by substituting  \eqref{4.30} into the expression for $k_i^{\theta, J}$ in  \eqref{4.10}.
From \eqref{4.30} we derive:
\begin{equation}\label{4.562}
 \frac{\int_{\mathbb{R}}e^{(\theta_i^{J,\ast} +1)x}\nu(dx)}{\int_{\mathbb{R}}e^{\theta_i^{J,\ast} x}\nu(dx)}=1
\end{equation}
As
\begin{equation}\label{4.563}
\int_{\mathbb{R}}e^{\theta_i^{J,\ast} x}\nu(dx)=\frac{1}{\sqrt{2\pi\sigma_J^2}}\int_{\mathbb{R}}e^{\theta_i^{J,\ast} x}e^{-\frac{(x-\mu_J)^2}{2\sigma_J^2}}dx=e^{\frac{1}{2}(\sigma_J \theta_i^{J,\ast})^2+\theta_i^{J,\ast}\mu_J}
\end{equation}
we obtain from \eqref{4.562} the following equality:
\begin{equation}\label{4.564}
e^{\frac{1}{2}(\sigma_J (\theta_i^{J,\ast}+1))^2+(\theta_i^{J,\ast}+1)\mu_J}=e^{\frac{1}{2}(\sigma_J \theta_i^{J,\ast})^2+\theta_i^{J,\ast}\mu_J}.
\end{equation}
The expression for the value of the  Esscher transform parameter $\theta_i^{J,\ast}$  in \eqref{4.56} follows immediately from \eqref{4.564}. Inserting this value of $\theta_i^{J,\ast}$ into the expression for $\lambda_i^{\theta, J}$ in \eqref{4.9} we obtain the formula \eqref{4.561}.

In the numerical simulations, we  assume that
the hidden Markov chain has three states: up, down, side-way, and the corresponding rate matrix is calculated using real Forex data for the thirteen-year period: from January 3, 2000 to November 2013. To calculate all probabilities we use the Matlab script (see the Appendix).

\section{Numerical simulations}
In the  Figures 2-4  we shall provide numerical simulations for the  case when the amplitude of jumps is described by a log-double exponential distribution. These three graphs show a dependence of the European-call option price against  $S/K$, where $S$  is the initial spot FX rate, $K$ is the strike FX rate for a different maturity time $T$ in years: 0.5, 1, 1.2. We use the following function in Matlab:
\begin{verbatim}Draw( S_0,T,approx_num,steps_num, teta_1,teta_2,p,mean_normal,sigma_normal)\end{verbatim}  to draw these graphs\footnote{Matlab scripts for all plots are available upon request}. The arguments of this function are: $S_0$ is the starting spot FX rate to define first point in $S/K$ ratio, \;$T$ is the maturity time, \;approx num describe the  number of attempts to calculate the mean for  the integral in the European call option pricing formula (see section 2, \eqref{4.39}), \;steps num denotes the  number of time subintervals to calculate the integral in \eqref{4.39};\;  teta 1, teta 2,\; p are $\theta_1, \theta_2, \;p$ parameters in the log-double-exponential distribution (see section 3, \eqref{4.40}). mean normal,\;sigma normal are the mean and deviation for the log-normal distribution (see section 4). In these three graphs: $\theta_1=10, \theta_2=10, p=0.5$; \; $\textnormal{mean normal}=0, \textnormal{sigma normal}=0.1$.

Blue line denotes the log-double exponential, green line denotes the log-normal,
red-line denotes the plot without jumps. The  $S/K$ ratio ranges from 0.8 to 1.25 with a step 0.05; the option price ranges from 0 to 1 with a step 0.1. The number of time intervals:  num =10.

From these three plots we conclude that it is important
to incorporate  jump risk into the spot FX rate models. Described by Black-Scholes equations without jumps,  red line on a plot is significantly below both blue and green lines which stand for the log-double exponential and the log-normal distributions of jumps, respectively.

All three plots have the same  mean value 0 and approximately equal deviations for both types of jumps: log-normal and log-double exponential. We investigate the case when it does not hold (see Figures 5-7).

As we can see, the log-double exponential curve moves  up in comparison with the log-normal and without jumps option prices.

If we fix  the  value of the $\theta_2$ parameter in the log-double exponential distribution with
 $S/K=1$  the corresponding  plot is given in Figure 8.

Figure 9 represents  a plot of the  dependence of the European-call option price against values of the parameters $\theta_1, \theta_2$ in a log-double exponential distribution, again  $S/K=1$.
\begin{figure}
\begin{center}
\includegraphics[ scale=0.5 ]{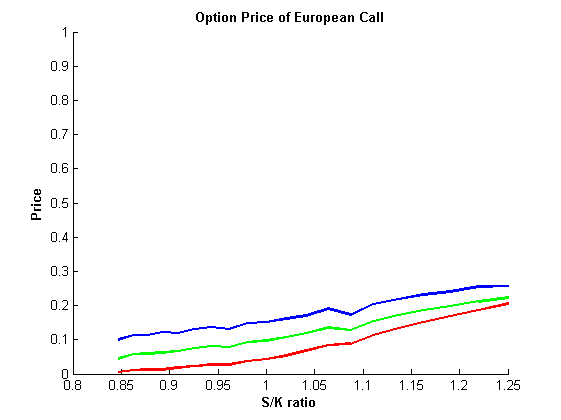}
\end{center}
\caption{$S_0=1, T=0.5, \theta_1=10, \theta_2=10, p=0.5, \textnormal{mean normal}=0, \textnormal{sigma normal}=0.1$}
\end{figure}
\begin{figure}
\begin{center}
\includegraphics[ scale=0.5]{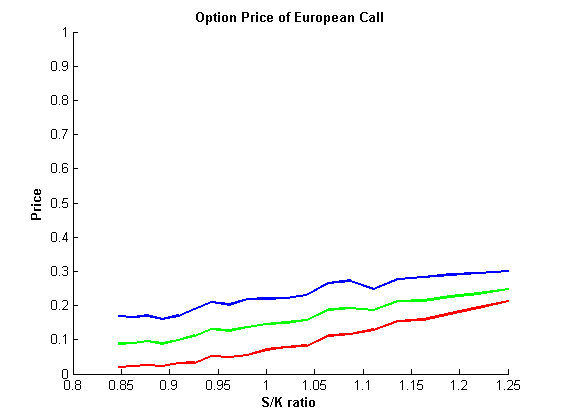}
\end{center}
\caption{$S_0=1, T=1.0, \theta_1=10, \theta_2=10, p=0.5, \textnormal{mean normal}=0, \textnormal{sigma normal}=0.1$}
\end{figure}
\begin{figure}
\begin{center}
\includegraphics[ scale=0.5]{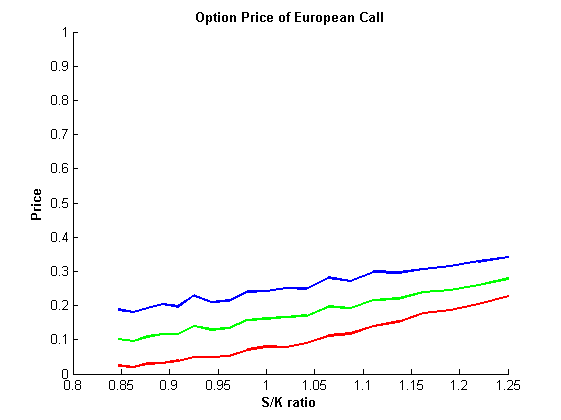}
\end{center}
\caption{$S_0=1, T=1.2, \theta_1=10, \theta_2=10, p=0.5, \textnormal{mean normal}=0, \textnormal{sigma normal}=0.1$}
\end{figure}
\begin{figure}
\begin{center}
\includegraphics[ scale=0.5]{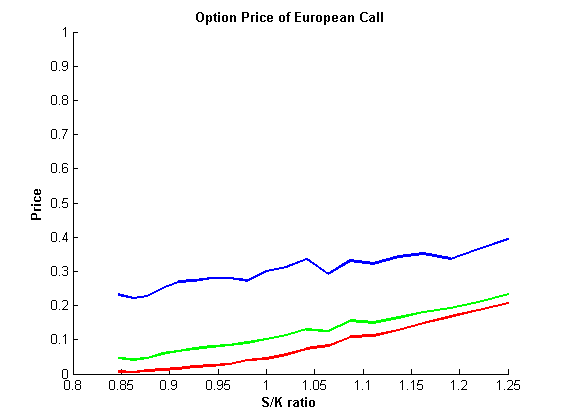}
\end{center}
\caption{$S_0=1, T=0.5, \theta_1=5, \theta_2=10, p=0.5, \textnormal{mean normal}=0, \textnormal{sigma normal}=0.1$}
\end{figure}
\begin{figure}
\begin{center}
\includegraphics[ scale=0.5]{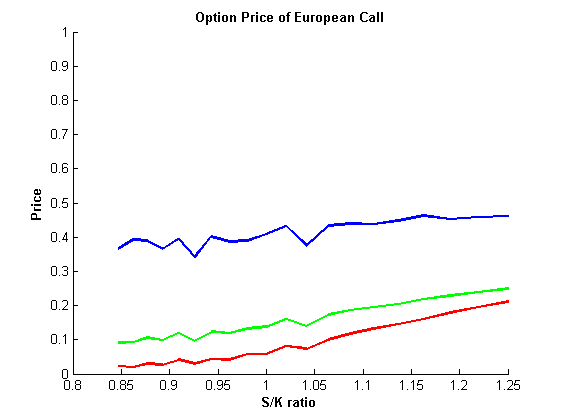}
\end{center}
\caption{$S_0=1, T=1.0, \theta_1=5, \theta_2=10, p=0.5, \textnormal{mean normal}=0, \textnormal{sigma normal}=0.1$}
\end{figure}
\begin{figure}
\begin{center}
\includegraphics[ scale=0.5]{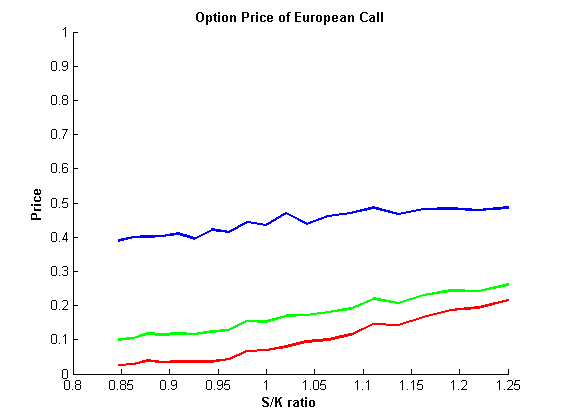}
\end{center}
\caption{$S_0=1, T=1.2, \theta_1=5, \theta_2=10, p=0.5, \textnormal{mean normal}=0, \textnormal{sigma normal}=0.1$}
\end{figure}
\begin{figure}
\begin{center}
\includegraphics[ scale=0.5]{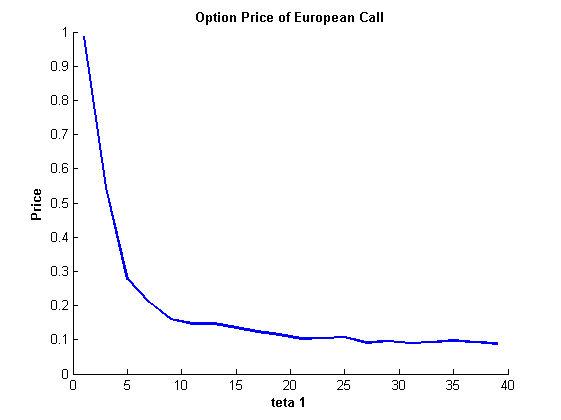}
\end{center}
\caption{$S_0=1, T=0.5, \theta_2=10, p=0.5$}
\end{figure}
\begin{figure}
\begin{center}
\includegraphics[ scale=0.5]{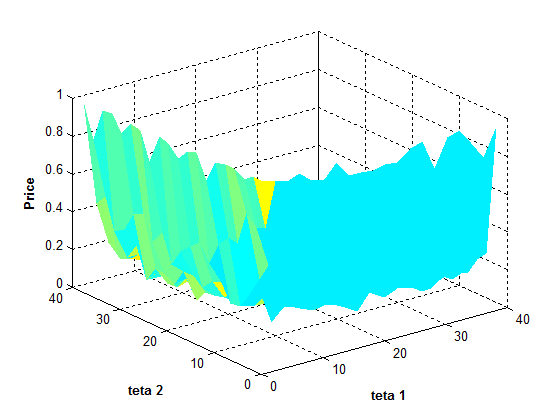}
\end{center}
\caption{Option price of European Call: $S_0=1, T=0.5,  p=0.5$}
\end{figure}

\newpage

\section{Conclusion}

In the conclusion, we generalized the results of  \cite{b7} to the case when the dynamics of the  FX rate is driven by a general L$\acute{\textnormal{e}}$vy process.  The main results of our research are as follows: 1) We generalized the formulas of  \cite{b7} for Esscher transform parameters which ensure that the martingale condition for the discounted foreign exchange rate is a martingale for a general L$\acute{\textnormal{e}}$vy process. Using the values of these parameters we proceeded to a risk-neural measure and provide new formulas for the distribution of jumps, the mean jump size, and  the Poisson process intensity with respect  to the measure.  Pricing formulas for European call foreign exchange options have been given as well (They are similar to those in \cite{b7}, but the mean jump size and the Poisson process intensity with respect to the new risk-neutral measure are different); 2) We applied obtained formulas to the case of  log double exponential processes; 3) We also provided numerical simulations of European call foreign exchange option prices for  different parameters. Codes for Matlab  functions used in numerical simulations of option prices are also provided.

\section*{Appendix}
\textbf{The Matlab function used to calculate probability matrix for the Markov chain modeling  cross rates of currency pairs in the Forex market.}

We assume that the Markov chain has only three states: "trend up", "trend down", "trend sideway". Such a choice of states is justified by numerous articles for the FX market (See www.mql5.com).
In a file $\textnormal{MaxDataFile open.CSV}$ there are open prices of EURO/ESD currency pairs of Japanese  candles over a 13 year period. This file was generated in the platform MT5 using MQL5 programming language.

\begin{verbatim}
function [ Probab_matrix ] =Probab_matrix_calc1(candles_back_up, candles_back_down,
delta_back_up, delta_back_down, candles_up,candles_down, delta_up, delta_down )
Probab_matrix=zeros(3,3);
m_open=csvread('MaxDataFile_open.CSV');
[size_open temp]=size(m_open);
m_before=zeros(1,size_open);
upper_border=size_open-max(candles_up, candles_down);
delta_up=delta_up/10000;
delta_down=delta_down/10000;
count_up=0;
count_down=0;
count_sideway=0;
beforeborder=max(candles_back_up, candles_back_down)+1;
for i=beforeborder:size_open
    if (m_open(i)-m_open(i-candles_back_up)>=delta_up)
        m_before(i)=1;
    end
     if (m_open(i-candles_back_down)-m_open(i)>=delta_down)
        m_before(i)=-1;
     end
end;
for i=1:upper_border
    if(m_before(i)==1)
        if(m_open(i+candles_up)-m_open(i)>=delta_up)
        Probab_matrix(1,1)= Probab_matrix(1,1)+1;
        else
            if(m_open(i)-m_open(i+candles_down)>=delta_down)
                Probab_matrix(1,2)= Probab_matrix(1,2)+1;
            else
                Probab_matrix(1,3)= Probab_matrix(1,3)+1;
            end
        end
    end
    if(m_before(i)==-1)
        if(m_open(i+candles_up)-m_open(i)>=delta_up)
        Probab_matrix(2,1)= Probab_matrix(2,1)+1;
        else
            if(m_open(i)-m_open(i+candles_down)>=delta_down)
                Probab_matrix(2,2)= Probab_matrix(2,2)+1;
            else
                Probab_matrix(2,3)= Probab_matrix(2,3)+1;
            end
        end
    end
    if(m_before(i)==0)
        if(m_open(i+candles_up)-m_open(i)>=delta_up)
        Probab_matrix(3,1)= Probab_matrix(3,1)+1;
        else
            if(m_open(i)-m_open(i+candles_down)>=delta_down)
                Probab_matrix(3,2)= Probab_matrix(3,2)+1;
            else
                Probab_matrix(3,3)= Probab_matrix(3,3)+1;
            end
        end
    end
end
count_up=sum(Probab_matrix(1,:));
count_down=sum(Probab_matrix(2,:));
count_sideway=sum(Probab_matrix(3,:));
for j=1:3
      Probab_matrix(1,j)= Probab_matrix(1,j)/count_up;
      Probab_matrix(2,j)= Probab_matrix(2,j)/count_down;
      Probab_matrix(3,j)= Probab_matrix(3,j)/count_sideway
end
end
\end{verbatim}

For example run in Matlab:
\begin{verbatim}
[ Probab_matrix ] = Probab_matrix_calc1(30, 30, 10, 10, 30, 30, 10, 10);
\end{verbatim}
Probability matrix is as follows:

$$
\left( {\begin{array}{cc}
    \textnormal{up} & \textnormal{down}\;\;\textnormal{sideway} \\
    0.4408 & 0.4527\;\; 0.1065\\
   0.4818 &0.4149 \;\; 0.1033\\
   0.4820 &0.4119\;\; 0.1061
  \end{array} } \right)
$$

\end{document}